\documentclass[twocolumn,preprintnumbers,amsmath,amssymb,superscriptaddress,prb]{revtex4}

\usepackage{graphicx}
\usepackage{dcolumn}
\usepackage{bm}
\usepackage{natbib}

\begin{document}

\title{Role of Diffusion in Two-dimensional Bimolecular Recombination}

\author {A. V. Nenashev}
\affiliation{Institute of Semiconductor Physics, 630090 Novosibirsk, Russia}
\affiliation{Novosibirsk State University, 630090 Novosibirsk, Russia}

\author {F. Jansson}
\affiliation{Graduate School of Materials Research,
\AA bo Akademi University,  20500 Turku, Finland}
\affiliation{Department of Physics and Center for Functional Materials,
\AA bo Akademi University, 20500 Turku, Finland}

\author {S. D. Baranovskii}
\affiliation{Department of Physics and Material Sciences Center,
Philipps-University, 35032 Marburg, Germany}

\author {R. \"Osterbacka}
\affiliation{Department of Physics and Center for Functional Materials,
\AA bo Akademi University, 20500 Turku, Finland}

\author {A. V. Dvurechenskii}
\affiliation{Institute of Semiconductor Physics, 630090 Novosibirsk, Russia}
\affiliation{Novosibirsk State University, 630090 Novosibirsk, Russia}

\author{F. Gebhard}
\affiliation{Department of Physics and Material Sciences Center,
Philipps-University, 35032 Marburg, Germany}

\date{\today}

\begin{abstract}
Experiments on carrier recombination in two-dimensional organic
structures are often interpreted in the frame of the Langevin model
with taking into account only the drift of the charge carriers in
their mutual electric field.  While this approach is well justified
for three-dimensional systems, it is in general not valid for
two-dimensional structures, where the contribution of diffusion can
play a dominant role.  We study the two-dimensional Langevin
recombination theoretically and find the critical concentration below
which diffusion cannot be neglected.  For typical experimental
conditions, neglecting the diffusion leads to an underestimation of
the recombination rate by several times.
\end{abstract}


\maketitle

The problem of charge carrier recombination is one of the central
problems in the solid state physics since recombination processes
determine all optoelectronic properties of a device or a material
under study. For instance, recombination can be an important factor
limiting the conversion efficiency in organic solar cells. In 
low-mobility organic semiconductors charge carrier recombination
is usually dominated by the Langevin bimolecular 
mode.\cite{Pope1999} 
Langevin recombination is characterized by the rate limiting step
being \textit{finding} the opposite charge carrier, independent of the
subsequent recombination mechanism.  It is well-known that in the
three-dimensional (3D) case the Langevin recombination rate is
determined by drift motion of charge carriers due to their mutual
Coulomb attraction and that diffusion can be
neglected.\cite{Greenham2003} The recombination rate per unit volume
equals $(\mu_e+\mu_h)eN_{e}^{2}/(\varepsilon \varepsilon_{0})$, where
$N_{e}$ is the electron density which is assumed to be equal to the
hole density $N_{e}=N_{h}\equiv N$, $\mu_e$ and $\mu_h$ are the
electron and hole mobilities, respectively, $\varepsilon
\varepsilon_{0}$ is the dielectric permittivity, and $e$ is the
electronic charge.\cite{Greenham2003,Juska2009,Juska2010} This result
is obtained by integrating the drift flux of electrons through a
spherical surface of radius $r$ surrounding a hole. Since the electric
field scales as $r^{-2}$ and the surface area of the sphere scales as
$r^2$, the value of $r$ chosen is unimportant, leading to a simple
solution with constant electron density, thus justifying the neglect
of diffusion.\cite{Greenham2003} The recombination rate per unit
volume is often written in the form $\beta N^{2}$ via the bimolecular
recombination coefficient $\beta = e (\mu_e+\mu_h)/\varepsilon
\varepsilon_0$, independent of carrier concentration in 3D.

Numerous recent experimental studies on bulk heterojunction solar cells have
demonstrated that the recombination in organic materials with
lamellar structures is much slower than what is predicted by
Langevin's theory.\cite{Pivrikas2005,Juska2005DI,Juska2006DI_Insul,Koster2006apl,Sliauzys2006tsf,Shuttle2008prb,Shuttle2008apl,Juska2008apl,Deibel2008apl,Lungenschmied2009}
Hence, in these films, where 2D conductivity mode is expected, the
bimolecular recombination coefficient depends on the
charge carrier density $N$.\cite{Juska2009,Juska2010,Shuttle2008apl,Juska2008apl,Deibel2008apl} 
The straightforward reformulation of the Langevin 3D
formalism for the 2D case indeed gives a concentration-dependent
recombination coefficient $\beta_{2D}\propto N^{1/2}$ if only the
drift motion is taken into account. This result was considered as an
explanation of the experimental data.\cite{Juska2009,Juska2010}

It is easy to understand the latter result rewriting the
recombination rate per unit volume in the form $\mathcal{R}_{rec}
= N/t$, where $t$ is the so-called recombination time. Remarkably
in any model that neglects diffusion, the recombination time in 2D is
proportional to $N^{-3/2}$, and consequently the recombination
rate per unit volume is proportional to $N^{5/2}$. It follows from
the fact that there is only one combination of the parameters
$\varepsilon_0\varepsilon$, $e$, $\mu$ and $N$ (not related to diffusion),
which has the dimensionality of time: $(\varepsilon_0\varepsilon/e\mu)N^{-3/2}$.

It is however not possible to neglect the contribution of
diffusion to the bimolecular Langevin recombination in 2D
systems.\cite{Greenham2003} In two dimensions, the electric field
still scales as $r^{-2}$, but the integration takes place over
a circle of circumference $2\pi r$. To achieve a recombination
current which is independent on $r$ requires a density of
electrons which varies with $r$. The presence of a carrier density
gradient means that diffusion must be explicitly included in the
problem.\cite{Greenham2003}

The combined effects of the drift and diffusion on the bimolecular
recombination in 2D has been studied by Greenham and
Bobbert.\cite{Greenham2003} In the following we first briefly repeat
their arguments. Afterwards we solve
the obtained equations analytically and compare the results with
Ref.~\onlinecite{Greenham2003}. Finally we reformulate the model of
Ref.~\onlinecite{Greenham2003} in order to make it compatible with the
model considered in Refs.~\onlinecite{Juska2009, Juska2010} in order to
estimate the effect of diffusion. It will be shown that the effect
depends essentially on the total concentration of charge carriers.

The following model was considered by Greenham and Bobbert.\cite{Greenham2003}
Electrons come uniformly to a
circular area around a hole. The radius $R$ of the
circle is chosen so that its inverse area is equal to the hole
concentration $N_h$:
\begin{equation} \label{eq-R-N}
(\pi R^2)^{-1} = N_h.
\end{equation}
The hole at the origin serves as a drain for incoming electrons.
Electrons can move within this area, due to drift in the electric
field of the hole,
\begin{equation} \label{eq-E}
\mathbf{E}(\mathbf{r}) = \frac{e}{4\pi\varepsilon_0\varepsilon} \,
\frac{\mathbf r}{r^3},
\end{equation}
and also due to diffusion. 

This model can be formalized via the following set of equations
for the electron flux density $\mathbf{J}(\mathbf{r})$ and the
coordinate-dependent electron concentration $n(\mathbf{r})$:
\begin{align} 
\nabla\cdot \mathbf{J}(\mathbf{r}) &= f, \label{eq-set1}\\
\mathbf{J}(\mathbf{r}) &= -\mu \, n(\mathbf{r}) \,
\mathbf{E}(\mathbf{r}) - D \, \nabla n(\mathbf{r}),\label{eq-set2}
\end{align}
that is fulfilled in the range $0<r<R$. Here $D$ is the diffusion 
coefficient ($D=\mu kT/e$ according
to Einstein's relation), $f$ is the incoming flux of electrons.
The boundary conditions for this problem are the following: $J(R)=0$
(there is no electron flux out of the circular area), and $n(0)=0$
(the hole is an ideal absorber of electrons). Solving this set of
equations, one can obtain the recombination time $t$ as
\begin{equation} \label{eq-t-n}
t = \langle n(\mathbf{r}) \rangle / f,
\end{equation}
where $\langle n(\mathbf{r}) \rangle$ denotes the concentration
averaged over the circular area. Then one can obtain the
recombination rate $\mathcal{R}_{rec}$ as
\begin{equation} \label{eq-Rrec}
\mathcal{R}_{rec} = N_e/t,
\end{equation}
where $N_e$ is the total electron concentration in the system.

The result of Greenham and Bobbert\cite{Greenham2003} can be
conveniently represented in terms of the dimensionless parameters
\begin{equation} \label{eq-tilde}
\tilde{R} = \frac R a, \qquad \tilde{t} = t \frac{D}{a^2},
\end{equation}
where $a=e^2/4\pi\varepsilon_0\varepsilon kT$ is the Onsager
radius:
\begin{equation} \label{eq-t-G}
\tilde{t} = \left(\tilde{R}^2-\frac12\right) \, G^{31}_{23}
\left(\frac{1}{\tilde{R}} \left|
\begin{array}{c}
0,3\\
0,0,2
\end{array} \right. \right) + \frac{\tilde{R}}{6} - \frac{\tilde{R}^2}{8},
\end{equation}
where G is the Meijer $G$ function. From
equations~(\ref{eq-R-N}), (\ref{eq-Rrec}), (\ref{eq-tilde}), and (\ref{eq-t-G})
one can obtain the concentration dependence of the recombination
time.

Greenham and Bobbert\cite{Greenham2003} analyzed the obtained
result numerically and suggested for the concentration dependence
of the recombination time a power-law interpolation: $t \sim
N_h^{-1.43}$. (As mentioned above, we will assume that electron
and hole concentrations are equal: $N_e=N_h\equiv N$.) Using
Eq.~(\ref{eq-Rrec}) one obtains then the following dependence for
the recombination rate: $\mathcal{R}_{rec} \sim N^{2.43}$.

One should however mention that the above interpretation of the
numerical result is valid only in a restricted range of
concentrations. In order to obtain a more general picture, let us
consider the limiting cases of large and small concentrations.
For this purpose let us use the following series expansions of the
Meijer $G$ function:
\begin{equation} \label{eq-G-expand1}
G^{31}_{23} \left(\frac{1}{x} \left|
\begin{array}{c}
0,3\\
0,0,2
\end{array} \right. \right) = \frac{x}{3}-\frac{x^2}{4}+\frac{2x^3}{5}+O(x^4) 
\text{ at } x\rightarrow0,
\end{equation}
\begin{equation} \label{eq-G-expand2}
G^{31}_{23} \left(\frac{1}{x} \left|
\begin{array}{c}
0,3\\
0,0,2
\end{array} \right. \right) = \frac{\log x}{2} -\frac{2\gamma+1}{4}+o(1) 
\text{ at } x\rightarrow\infty.
\end{equation}
Here $\gamma\approx0.577$ is Euler's constant.

Let us first consider the limit of large concentrations, i.~e.
small $\tilde{R}$. Taking into account the
expansion~(\ref{eq-G-expand1}), it is easy to show that
\begin{equation} \label{eq-tilde-t-large-c}
\tilde{t} =  \frac{2}{15}\, \tilde{R}^3 + O(\tilde{R}^4),
\end{equation}
whence the dependence $t(N)$ follows the relation:
\begin{equation} \label{eq-t-large-c}
t =  \frac{2}{15\pi^{3/2}aD}\, N^{-3/2} \equiv
\frac{8}{15\sqrt{\pi}} \, \frac{\varepsilon_0\varepsilon}{e\mu} \,
N^{-3/2}.
\end{equation}
Consequently, $\mathcal{R}_{rec} \sim N^{2.5}$, in accordance to
the result of Ju\v{s}ka \emph{et al.},\cite{Juska2009} where only
drift was taken into account and diffusion was neglected.
Remarkably, \emph{in the case of high carrier concentrations $N$
the contribution of diffusion to the recombination rate is
not important}. Notably our coefficient of proportionality
($15\sqrt{\pi}e\mu/8\varepsilon_0\varepsilon$) appears 2.5 times
larger than the one given in Ref.~\onlinecite{Juska2009}. This
difference comes from different assumptions in the models: in
Ref.~\onlinecite{Juska2009} each electron was considered to start
at the same distance $R$ from the hole, while in the model considered above
(Ref.~\onlinecite{Greenham2003}), the
starting points of electrons were chosen randomly within the
circle of radius $R$.

\begin{figure}
\includegraphics[width=8.6 cm]{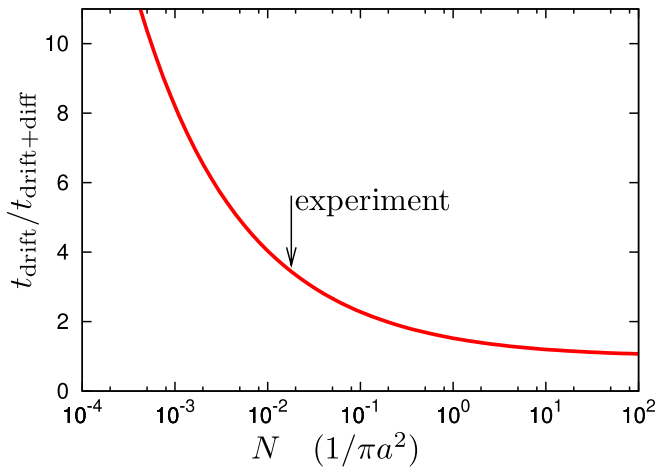}
\caption{The ratio between the recombination times $t_\text{drift}$
  obtained by taking only drift into account
  [Eq.(\ref{eq-t-large-c-m})] and $t_\text{drift+diff}$ obtained when
  considering both drift and diffusion [Eq.(\ref{eq-t-G-m})], as a
  function of the charge carrier concentration $N$.  The concentration
  is measured in units of the number of charge carriers inside the
  Onsager radius $a$.  The arrow corresponds to the experimental
  concentration in Ref.~\onlinecite{Juska2009}.}
\label{fig-1}
\end{figure}

We reformulate
the formalism of Greenham and Bobbert\cite{Greenham2003}
adjusting it to the model assumptions of Ju\v{s}ka \emph{et
al.}\cite{Juska2009, Juska2010} To do so one has to replace the
uniform influx of electrons inside the circular area by the influx
from only the periphery of the circle. Then Eq.~(\ref{eq-set1})
takes the form:
\begin{equation} \label{eq-set1-m}
\nabla\cdot \mathbf{J}(\mathbf{r}) = 0,
\end{equation}
with replacing the boundary condition $J(R)=0$ by the relation
$J_z(R) = -fR/2.$
This condition provides the same average influx of electrons as
Eq.~(\ref{eq-set1}).
The solution of this modified problem gives instead of
Eq.~(\ref{eq-t-G}) the following expression:
\begin{equation} \label{eq-t-G-m}
\tilde{t} = \tilde{R}^2 \, G^{31}_{23} \left(\frac{1}{\tilde{R}}
\left|
\begin{array}{c}
0,3\\
0,0,2
\end{array} \right. \right) .
\end{equation}
In the limit of small $\tilde{R}$ (i.~e. high concentrations) one
can keep only the first term of the expansion~(\ref{eq-G-expand1})
and obtain:
\begin{equation} \label{eq-tilde-t-large-c-m}
\tilde{t} =  \frac{1}{3}\, \tilde{R}^3 + O(\tilde{R}^4),
\end{equation}
and consequently
\begin{equation} \label{eq-t-large-c-m}
t = \frac{4}{3\sqrt{\pi}} \, \frac{\varepsilon_0\varepsilon}{e\mu}
\, N^{-3/2},
\end{equation}
which is exactly the result of Ju\v{s}ka \emph{et
al.}\cite{Juska2009, Juska2010}
Figure \ref{fig-1} illustrates that at high carrier concentrations one
can indeed neglect the contribution of diffusion to the recombination time.
However, the smaller is the concentration the more important is the
contribution of the diffusion process. For instance, at carrier
concentrations in the experiments reported in
Ref.~\onlinecite{Juska2009} neglecting the contribution of
diffusion gives a recombination time three times longer than when diffusion is included. 
At lower concentrations this overestimation can reach orders
of magnitude.  Fig.~\ref{fig-1} shows that within the accuracy of 50
\% the neglect of diffusion is justified for concentrations $N \gtrsim
N_c = \frac {1}{\pi a^2}$.  In other words, if the average distance
between an electron and a hole is larger than the Onsager radius $a$,
which is the case at low charge carrier concentrations,
one cannot neglect the contribution of diffusion to the Langevin
recombination in 2D systems.

Let us therefore pay more attention to the limit of low concentrations.
Using equations (\ref{eq-R-N}), (\ref{eq-tilde}), 
(\ref{eq-t-G}), (\ref{eq-G-expand2}), and 
(\ref{eq-t-G-m}) one comes to the conclusion that in this case the
concentration dependence of the recombination time has the 
form:
\begin{equation} 
t = \frac{1}{4\pi D}\, N^{-1} \log\frac{\alpha}{a^2N},
\label{eq-t-small-c}
\end{equation}
where the numeric constant $\alpha$ is $1/(\pi
e^{2\gamma+3/2})\approx0.022$ for uniform generation of carriers
inside the circular area, as considered in
Ref.~\onlinecite{Greenham2003}, and $1/(\pi
e^{2\gamma+1})\approx0.037$ for generation of carriers on the
periphery of the circle, as considered in Refs.~\onlinecite
{Juska2009, Juska2010}.  In the limit $N\rightarrow0$, the
recombination rate reads
\begin{equation} \label{eq-rate-small-c}
\mathcal{R}_{rec} = \frac{N}{t} = 4\pi D\, N^2 
\left( \log\frac{\alpha}{a^2N} \right)^{-1}.
\end{equation}
Using for the recombination rate per unit volume the traditional
notation $\beta N^2$ we conclude that in the limit of low
concentrations $N$, the bimolecular recombination coefficient $\beta$
depends logarithmically on $N$. This dependency is essentially weaker
than the one predicted by considering solely the drift
processes.\cite{Juska2009,Juska2010}

We are indebted to Prof.\ Gytis Ju\v{s}ka for stimulating discussions
and to Dr. A. Zinovieva for reading the manuscript and giving valuable
comments.  Financial support from the Academy of Finland project
116995, from the Deutsche Forschungsgemeinschaft and from the Fonds
der Che\-mischen Industrie is gratefully acknowledged.

\end{document}